\documentstyle[twoside,fleqn,espcrc2]{article}

\def\a{\alpha}

\def\d{\delta}
\def\f{\phi}                    

\def\k{\kappa}

\def\m{\mu}
\def\n{\nu}

\def\th{\theta}                  

\def\D{\Delta}

\def\cd{{\cal D}}

\def\cl{{\cal L}}

\def\bo{\raisebox{-.4ex}{\large$\Box$}}                 
\def\cbo{{\,\raise-.15ex\Sc [\,}}                       
 
\def\svev#1{\left\langle #1\right\rangle}       

\def\ddt#1{{\buildrel {\hbox{\LARGE .\kern-2pt.}} \over {#1}}}

\def\beq{\begin{equation}}
\def\eeq{\end{equation}}
\def\bqry{\begin{eqnarray}}
\def\eqry{\end{eqnarray}}

\def\beqn#1{ \renewcommand{\theequation}{#1} 
             \begin{eqnarray} }
\def\eeqn{ \renewcommand{\theequation}{\arabic{equation}}
           \end{eqnarray} }

\def\beqr#1{ \setcounter{equation}{#1} 
             \begin{eqnarray} }

\def\eeqr{\end{eqnarray}}
\def\NON{\nonumber\\}
\def\beqrabc#1{ \setcounter{equation}{0}
                \renewcommand{\theequation}{#1\alph{equation}} 
                \begin{eqnarray} }
\def\beqrn#1#2{ \setcounter{equation}{#2}
                \renewcommand{\theequation}{#1.\arabic{equation}} 
                \begin{eqnarray} }
\def\seeq#1{eq.~(\ref{#1})}

\def\rf{ref.~\cite}

\def\NPB#1{Nucl. Phys. {\bf B#1}}
\def\NPBP#1{Nucl. Phys. (Proc. Suppl.) {\bf B#1}}
\def\PLB#1{Phys. Lett. {\bf B#1}}

\def\PRL#1{Phys. Rev. Lett. {\bf #1}}

\def\sstyle{\scriptstyle}

\def\frac#1#2{ {\sstyle {#1\over #2} } }

\def\tk{\tilde\k}

\newcommand{\AmS}{{\protect\the\textfont2
  A\kern-.1667em\lower.5ex\hbox{M}\kern-.125emS}}

\title{Lattice Chiral Gauge Theories in a Renormalizable Gauge}

\author{Yigal Shamir\address{School of Physics and Astronomy,
Beverly and Raymond Sackler Faculty of Exact Sciences,\\
Tel-Aviv University, Ramat Aviv 69978, ISRAEL}%
        \thanks{Work supported in part by the US-Israel Binational Science 
Foundation, and the Israel Academy of Science.}}

\begin{document}

\begin{abstract}
The lattice formulation of gauge theories in a renormalizable
gauge is discussed. The formulation invokes a new 
phase diagram, and it may allow for a
lattice definition of Chiral Gauge Theories.
\end{abstract}

\maketitle

\section{INTRODUCTION}

  We report here on a new approach~\cite{sml} to the long-standing problem
of constructing chiral gauge theories using the lattice regularization.
A renormalizable gauge fixed action plays a central role in the
lattice formulation. It was first suggested in \rf{roma}
that gauge fixing is vital for the construction of Lattice Chiral 
Gauge Theories (LCGTs). Here we make an important step forward,
and explicitly construct a lattice model where the gauge fixing paradigm
is realized.

  Most of the modern attempts to define LCGTs invoke a fermion action where
gauge invariance is explicitly broken, thus avoiding a direct conflict with
well-known No-Go theorems~\cite{nogo}. In a gauge invariant theory 
like QCD, the fermion spectrum can be read off from the lattice action 
by going to the free field limit. But in the absence of exact gauge invariance,
the degrees of freedom along the lattice gauge orbit
{\it couple to the fermions}, and one has to understand the
consequences of this coupling.

  Under these circumstances, the fermion spectrum is determined 
by a {\it reduced model}, in which only the 
longitudinal modes of the lattice gauge field are kept. 
The action of the reduced model is obtained by substituting 
$U_{x,\m} = \f_x \f^\dagger_{x+\hat\m}$ into the original action,
where $\f_x\in G$ is a group valued scalar field.
The original measure $\int \cd U$  is replaced by $\int D\f$. 
Formally, the reduced model corresponds to
setting $g_0=0$ in the action. Thus, finding the fermion spectrum by
going to the reduced model is a natural generalization of
what one does in the gauge invariant case. In both cases,
the procedure is justified because the continuum limit corresponds to
$g_0 \to 0$.

  One can go back from the reduced model to the original model
in two steps. In the first step one
gauges the reduced model. This leads to a manifestly gauge
invariant, generalized Higgs model, where both $\f_x$ and $U_{x,\m}$
appear as independent fields (``generalized'' means that 
the higgs action is not the conventional one). In the second step,
gauge invariance of the generalized Higgs model is used 
to completely eliminate the $\f_x$ field, while leaving
the partition function invariant~\cite{phi}. Any observable defined from
the {\it vector picture} of the theory (where the partition function 
contains only the $U_{x,\m}$ field), coincides with a corresponding 
gauge invariant observable defined from the partition function in the 
{\it Higgs picture}. The group valued field is seen to play the dual
role of a Higgs or a St\"uckelberg field. Which interpretation captures 
the physics better depends on the {\it dynamics}.

  The original gauge group reappears as an exact 
global symmetry of the reduced model, that
acts on the Higgs--St\"uckelberg field $\f_x$ by {\it right multiplication}. 
This global symmetry serves to assign the fermions (or any other matter fields)
to representations of the gauge group. Thus,
one can map out the phase diagram, and study the fermion spectrum in
the various phases.

  In most models that have been investigated so far, the fluctuations 
of the Higgs--St\"uckelberg field are not controlled 
by any small parameter. This has the consequence that one cannot rely on
perturbation theory for finding the fermion spectrum. 
Where non-perturbative methods are available, 
one finds that either the fermion spectrum is 
vector-like (symmetric phase) or the gauge bosons have acquired cutoff masses
and decoupled (broken phase). The fact that the fermion spectrum is 
vector-like in symmetric phases can be understood in terms
of a generalized No-Go theorem~\cite{gennogo}. For a more detailed discussion
including references to the original literature see \rf{rev}.

\section{A NEW PHASE DIAGRAM}

  The attempts to construct LCGTs in a symmetric phase lead to an impasse.
This raises the question, can we do better in the broken phase?
In the broken phase the doublers can acquire a mass $m_d \sim y v$,
where $v$ is the Higgs VEV and $y$ is a Yukawa (or Yukawa-Wilson)
coupling. (In the non-abelian case we assume a matrix valued VEV
$v_{AB} = v \d_{AB}$.)
Keeping the primary fermions massless is not a problem~\cite{sml}.
Now, if we want to decouple the doublers, we have to send $v$ to infinity, 
in physical units. (The formula $m_d \sim y v$ holds only for small $y$. 
There is conclusive evidence that the fermion spectrum is different in
the limit $y \to \infty$, and that one is back to the same problem as in
the weak coupling symmetric phase).
However, for a conventional Higgs theory, gauge invariance implies
the relation $m \sim gv$ for the {\it gauge bosons} mass. Sending
$v$ to infinity will then decouple the gauge bosons too, and we seem
to have ended in a different impasse.

  What happens if we allow for a Higher Derivative (HD)
Higgs action? Now the formula $m \sim gv$ is no longer valid.
In the classical approximation, a HD Higgs action gives rise to no gauge 
boson mass in the broken phase. As a result, one can send $v$ to infinity
and decouple the doublers, while retaining the gauge bosons in
the low energy spectrum. We use this loophole
to escape from the previous impasses. 

  A HD action will eventually give rise to {\it some} gauge boson mass via
quantum corrections. Therefore, we must be able to {\it tune} the renormalized 
vector boson mass  $m_r$ to zero, while staying in the broken phase.
If we have a parameter that allows us to adjust the curvature
of the potential of a vector field, and tune it to zero at the origin, then 
there should be a parameter range where the curvature at the origin 
is {\it negative}. In this range condensation of a {\it vector field} 
will take place.

  The phase structure that we need  thus consists of 
two broken phases: an ordinary broken phase and a new one, denoted FMD, which 
has a preferred {\it direction} defined by a vectorial order parameter. 
In the reduced model, the vectorial order parameter corresponds to
a non-zero momentum of the ferromagnetic ground state, and the  
phase transition separates the FMD phase from an FM phase.
For $g_0 \ne 0$, a vector field develops a non-zero VEV in the FMD phase.
(There are no physical Goldstone modes for $g_0 \ne 0$, because the 
lattice rotation group is discrete.) With ``ordinary broken phase,''
we refer to the rotationally invariant region on the other side of 
the phase boundary, which belongs to a Higgs or Higgs-confinement phase.

  The continuum limit is defined by approaching a gaussian critical point
on the FMD phase boundary (see below). This new continuum limit 
is qualitatively different from the conventional Higgs transition. 
The cutoff masses acquired by the fermion doublers are attributed 
to the scalar VEV in the reduced model. In this sense, 
$\f_x$ plays the usual role of a {\it Higgs field}
from the fermion's point of view. (The primary fermions remain massless
because their Yukawa couplings are set to zero.) From the point of view 
of the gauge sector, however, $\f_x$ plays the role of a 
{\it St\"uckelberg field}. The reason is that the continuum
limit is taken far from any symmetric phase to keep $v\sim 1$ 
in lattice units. As a result, radial fluctuations of the $\f_x$ field 
(which do not exist classically) are suppressed also at the quantum level.

  A non-trivial continuum limit should be described by some renormalizable
continuum lagrangian. In the absence of a Higgs resonance,
the new critical region is controlled by a {\it renormalizable vector 
lagrangian}, which can be read off from the lattice action by
going to the {\it vector picture} of the full model. 
This vector lagrangian contains kinetic terms for {\it all} 
polarizations. A transversal kinetic term is provided as usual by the 
plaquette term, whereas a longitudinal kinetic term
$(\partial \!\cdot\! A)^2$ arises naturally if
one chooses the simplest HD Higgs action. We believe that the 
longitudinal kinetic term is in fact indispensable, and that an attempt
to get rid of it by some tuning is bound to end up in  
uncontrollable IR divergences. 

  A renormalizable, but otherwise arbitrary, vector theory is not unitary.
In view of the presence of a longitudinal kinetic term, 
the idea is to bring the marginal gauge
symmetry breaking terms in the vector lagrangian to the form
\beq
  {1\over 2 \a_0}\, (\mbox{gauge condition})^2 \,.
\label{gcond}
\eeq
This allows the interpretation of a {\it gauge fixing} action.
The gauge fixing term can be either $(\partial \!\cdot\! A)^2$ \cite{gfx} or,
alternatively, $(\partial \!\cdot\! A + g A^2)^2$ \cite{sml}, which corresponds 
to a non-linear gauge. An appropriate Faddeev-Popov (FP) ghost action 
has to be included. Provided the fermion spectrum is anomaly free, 
perturbative unitarity is now recovered by adding
appropriate counter-terms to enforce 
the Slavnov-Taylor (BRST) identities in the continuum limit,
as first proposed in \rf{roma}. The issue of exact unitarity is left open. 
Notice that one of the BRST identities is $m_r=0$. This BRST identity is at
the heart of our approach, as it amounts to approaching
the FMD phase boundary in the continuum limit.

  All the new features of our approach have to do with the treatment of
the gauge sector of the theory. A success in achieving the critical 
behaviour described above, will allow us to couple the gauge field
to chiral fermions using one of several existing fermion actions.
Thus, what we have done boils down to proposing an answer to the question 
of how to define lattice gauge theories  in a {\it renormalizable gauge}.
It is interesting that the qualitative features of the desired phase diagram
can be deduced by requiring consistency of the gauge fixing procedure, 
while making no reference to the presence or absence of fermions. 
This alternative point of view is discussed in \rf{gfx}.

\section{THE FORMULATION}

We first present a simple non-linear model that provides 
the basic phase diagram that we need. 
The vector theory that gives rise to this reduced model 
can be reconstructed as described in the introduction.  
The action, which borrows from previous work on 
higher derivative models~\cite{jkl}, is given by 
\beq
  S_H =  \sum_{xy} \left( -\k\, \f^\dagger_x \bo_{xy} \f_y 
  + \tk\, \f^\dagger_x \bo^2_{xy} \f_y \right),
\label{sh}
\eeq
where $\bo_{xy}$ is the standard nearest-neighbor laplacian. The lattice
spacing $a$ is set equal to one. 
The $\f_x$ field takes values in some compact Lie group. For simplicity
we will consider the case $\f_x \in {\rm U(1)}$. The generalization
to non-abelian theories is given in \rf{sml,gfx}. 

  As mentioned earlier, the vectorial order parameter of the non-linear model
takes the form of a non-zero momentum for the ground state. Assuming 
\beq
  \svev{\f_x} = v e^{iqx} \,,
\label{vq}
\eeq
one can study the phase diagram using standard mean-field
techniques. For large $\tk$ one is in a broken phase and $v\ne 0$.
Minimizing the mean-field hamiltonian with respect to $q_\m$, one finds
a second order transition between an FM phase and an FMD phase at $\k=0$, 
with $q_\m\ne 0$ for $\k<0$. 
(In the vector picture of the full model, the classical vector VEV is 
$g_0\svev{A_\m}=q_\m$.)

  A central question is whether the continuity of the FMD transition 
is not too much spoiled by quantum effects. In the reduced model,
quantum corrections can be studied in a {\it systematic 
expansion} around $\tk=\infty$, which is a zero temperature limit.  
The critical region 
near the FMD phase boundary is governed by a Goldstone Boson lagrangian
whose coupling constant is $1/\tk$. The GB lagrangian is derived
by substituting $\f_x = \exp(i\th_x/\sqrt{2\tk})$ into the action.
$\th_x$ is the Goldstone field. The classical continuum limit of the
lattice action leads to
\beq
  \cl_{GB} = {\k\over 2\tk}\, \partial_\m\th\, \partial_\m\th
	     + {1\over 2}\, (\bo\th)^2 
             + {1\over 4\tk}\, (\partial_\m\th\, \partial_\m\th)^2 .
\label{gb}
\eeq
This lagrangian exhibits the unusual feature of containing a $p^4$
kinetic term. (The analytic continuation of the GB lagrangian to Minkowski
space is not unitary, but this is of no direct concern to us, since
unitarity in the continuum limit is needed only after the
gauge field is introduced). The GB lagrangian is renormalizable, and
leads to correlation functions of the $\partial_\m\th$ field obeying the
standard power counting both in the UV and the IR limits. The situation 
with respect to correlation functions of the $\f$-field is more subtle.

Gauging the reduced model amounts to replacing the ordinary laplacian with 
a covariant one in \seeq{sh}, and adding the usual plaquette action for 
the gauge field. After going to the vector picture (where $\f_x=1$),
we find that the covariant HD action leads to the following new terms
\bqry
  \tk \left. \f^\dagger \bo^2(U) \f  \right|_{\f_x=1}  = \mbox{\hspace{4cm}}  
\NON 
  {1\over 2\a_0} \Big( \sum_\m \D^-_\m A_\m \Big)^2
  + {g_0^2\over 2\a_0} \Big( \sum_\m A^2_\m \Big)^2 + \cdots  .
\label{exp}
\eqry
Here we have defined $1/2\a_0=\tk g_0^2$, and $\D^-_\m$ is the backward 
lattice derivative. We made use of the standard weak coupling expansion
$U_\m = \exp( i g_0 A_\m )$. The dots stand for irrelevant terms. 
This result is still not acceptable, however. What we need is an additional 
term in the HD Higgs action that does not spoil the phase diagram, 
whose effect is to bring the vector action to the form of some gauge 
fixing action, {\it cf.} \seeq{gcond}. For the linear gauge 
one has to cancel the $(\sum_\m A^2_\m )^2$ term, whereas for the 
non-linear gauge one has to add a mixed term proportional to
$(\sum_\m \D^-_\m A_\m) (\sum_\n A^2_\n)$. A detailed construction of
the complete HD action is given in \rf{gfx}. In both cases, the
marginal terms in the lattice gauge fixing action are
lattice transcriptions of the corresponding continuum expressions.
The irrelevant terms are chosen such that the 
unique minimum of the gauge fixing action is $U_{x,\m}=I$.
Consequently, {\it lattice artefact Gribov copies} are
suppressed, and perturbation theory is a valid starting point for
the investigation of the model.

  As can be seen from \seeq{exp}, it is natural to take $1/\tk$ to scale like
$g_0^2$. In the continuum limit one therefore approaches the {\it gaussian 
critical point} $g_0=1/\tk=0$. For $g_0=0$, we expect that the FMD 
transition (as a function of the scaling variable $\k/\tk$) will remain
continuous in the limit $\tk\to \infty$. Off the critical point,
in particular for $g_0 \ne 0$, the transition may become weakly first
order. In view of the classical stability of the FMD transition,
and the consistency of the weak coupling expansion in $g_0$ and
$1/\tk$, any dynamically generated IR scale should be
a non-perturbative function of the coupling constants. This, in turn,
should imply the existence of a scaling region.

  In conclusion, the coupling between the fermions and the gauge degrees 
of freedom entails the need for a good control over the longitudinal
dynamics. This can be achieved by formulating lattice gauge theories
in a renormalizable gauge. The generic phase diagram needed for this
new formulation was discussed above.
Promoting the standard perturbative gauge fixing procedure to a 
non-perturbative one requires us to study many new questions.
Work on a number of issues is in progress.


\begin{thebibliography}{9}

\bibitem{sml} Y.\ Shamir, hep-lat/9512019 (revised May 1996).

\bibitem{roma} A.\ Borelli, L.\ Maiani, G.-C.\ Rossi, 
R.\ Sisto and M.\ Testa, \PLB{221} (1989) 360; \NPB{333} (1990) 335.

\bibitem{nogo} L.H.\ Karsten and J.\ Smit, \NPB{183} (1981) 103;
H.B.\ Nielsen and M.\ Ninomiya, \NPB{185} (1981) 20, {\bf B193} 
(1981) 173; {\it Erratum}, \NPB{195} (1982) 541.

\bibitem{phi} D.\ Foerster, H.B.\ Nielsen and M.\ Ninomiya, 
\PLB{94} (1980) 135; J.\ Smit, \NPBP{4} (1988) 451;
S.\ Aoki, \PRL{60} (1988) 2109;
K.\ Funakubo and T.\ Kashiwa, \PRL{60} (1988) 2113.

\bibitem{gennogo} Y. Shamir, \PRL{71} (1993) 2691; \NPBP{34} (1994) 590;
hep-lat/9307002.

\bibitem{rev} Y.\ Shamir, Plenary talk at Lattice'95, Melbourne, Australia,
\NPBP{47} (1996) 212.

\bibitem{gfx} M.F.L.\ Golterman and Y.\ Shamir, hep-lat/9608116.

\bibitem{jkl} K.\ Jansen, J.\ Kuti and C.\ Liu, \PLB{309} (1993) 119, 127;
\NPBP{30} (1993) 681, {\bf B34} (1994) 635, {\bf B42} (1995) 630;
J.\ Kuti, \NPBP{42} (1995) 113.

\end{thebibliography}
\end{document}